\documentclass{article}

\usepackage{arxiv}

\usepackage[utf8]{inputenc} 
\usepackage[T1]{fontenc}    
\usepackage{hyperref}       
\usepackage{url}            
\usepackage{booktabs}       
\usepackage{amsfonts}       
\usepackage{nicefrac}       
\usepackage{microtype}      
\usepackage{lipsum}

\usepackage{amssymb,amsfonts,amsbsy,amsmath}
\usepackage{subfigure}
\usepackage{comment}
\usepackage{xparse}
\usepackage{enumitem}
\usepackage{multirow}
\usepackage{tikz,pgfplots}
\usetikzlibrary{shapes,arrows,shadows,positioning}
\usetikzlibrary{matrix,chains,positioning,decorations.pathreplacing}

\usepackage{graphicx,xcolor}
\usepackage{subfigure}
\usepackage{epstopdf} 
\usepackage[export]{adjustbox} 
\usepackage{adjustbox} 
\usepackage{longtable} 

\usepackage[]{algorithm2e,algpseudocode}

\DeclareMathOperator*{\argmax}{arg\,max}

\newcommand{\bs}[1]{\boldsymbol{#1}}

\title{A quadratic linear-parabolic model-based classification to detect epileptic EEG seizures}
		
\author{
  Antonio Quintero-Rinc\'on, Carlos D'Giano \\
  Fundaci\'on Lucha contra las Enfermedades Neurol\'ogicas Infantiles (FLENI)\\
  Buenos Aires, Argentina \\
  \texttt{tonioquintero@ieee.org} \\
   \And
 Hadj Batatia \\
  University of Toulouse, IRIT - INPT\\
	Toulouse, France}
		
\begin{document}
\maketitle
		
		\begin{abstract}
			The two-point central difference is a common algorithm in biological signal processing and is particularly useful in analyzing physiological signals. In this paper, we develop a model-based classification method to detect epileptic seizures that relies on this algorithm to filter EEG signals. The underlying idea is to design an EEG filter that enhances the waveform of epileptic signals. The filtered signal is fitted to a quadratic linear-parabolic model using the curve fitting technique. The model fitting is assessed using four statistical parameters, which are used as classification features with a random forest algorithm to discriminate seizure and non-seizure events. The proposed method was applied to 66 epochs from the Children Hospital Boston database. Results show that the method achieves fast and accurate detection of epileptic seizures, with a 92\% sensitivity, 96\% specificity, and 94.1\% accuracy.
			
		\end{abstract}
		
\keywords{Two-point central difference \and curve fitting \and parabolic curves \and Epilepsy\and EEG \and Random forest}

	\section{Introduction}
	Epilepsy is a neurological disease that affects people of all ages. It is characterized by unpredictable seizures resulting from the hyperexcitability of neurons. The electroencephalogram (EEG) is the predominant modality to study cerebral activity.  The analysis of EEG signals is nearly completely dependent on visual inspection by the physician to quantify or qualify the morphology of waves, their goal being the identification and classification of  abnormal patterns in order to provide aid for an epilepsy diagnosis.
	
	This study proposes a simple, fast and adaptable method, implementable in real-time, to help physicians visually inspect EEG signals for epileptic seizure detection. 
		The proposed method fits within the framework of model-based classification. It is based on the statistical parameters obtained from fitting a quadratic parabolic model to \textit{specifically} filtered and transformed EEG signals. Precisely, the two-point central difference algorithm is used to build a filter for the EEG signals. The filtered signal is subsequently represented using a quadratic linear-parabolic model, using the curve fitting approach. Four statistical parameters are then calculated to characterize the model fitting. These parameters are considered as a feature-vector in the classification stage.
	
	The literature abounds with research work dealing with the detection of epileptic seizure onset. Various methods and techniques have been proposed for this purpose. Many existing methods fit within the large framework of the feature-based machine learning approach.
	
	The two-point central difference algorithm has proven useful in analyzing physiological signals, with various medical applications reported in the literature. A typical work has been reported by Frei et al. to detect epileptiform discharges in muscles in EEG signals \cite{Frei1999}. This algorithm is usually used to estimate the derivative of a function, giving an estimate valid only over a limited frequency range. Here, our idea is to use this algorithm to design a filter that enhances the frequency range of EEG signals corresponding to brain activity characteristic of epileptic seizures. To single out the related waveform, we transform the filtered signal into a quadratic form. A quadratic linear-parabolic model is then fitted to the transformed signal, using the curve fitting technique \cite{Chernov2004,LopezRubio2018}. The fitting is assessed using four statistical parameters (weighted sum of squared residuals, R-square, adjusted R-square, and root mean squared error). In order to show that our quadratic linear-parabolic model is pertinent to characterize epileptic seizures, we develop a classification method where the model-fitting parameters are used as features. We show that these features are good bio-markers of epileptic seizures. 
	
	The curve fitting technique adopted here is widely used in signal processing in general and in biomedical applications in particular. It has also been  applied a few times in the domain of EEG and epilepsy. Orhan et al. used the polynomial curve fitting method to fit a probability density function (PDF) of EEG signals discretized using equal frequency binning. The method has been applied to epileptic seizure detection using the comparison of PDFs of seizures and non-seizures. In \cite{Zhang2018a}, second-order Fourier curve fitting has been used to smooth EEG signals in a pre-processing stage of a model-based method to predict epileptic seizures. A spline curve fitting to EEG data was proposed in \cite{Ramon2018} to detect phase cone patterns to characterize epileptic activity.
		In \cite{Liu2018}, linear curve fitting was applied to interictal heartbeat intervals in order to combine EEG and ECG signals with the purpose of characterizing modulation patterns in patients with drug-resistant epilepsy.
		More recently, a first-order exponential function has been fit to voltage-gated sodium channel type 2 to model the conductance-voltage curves in neonatal epileptic subjects \cite{Begemann2019}.
	
	Supervised machine learning methods have widely been used to detect seizures in EEG, with support vector machine (SVM) and K-Nearest Neighbor (KNN) being the most popular. SVM classifiers have been used with various features extracted from spectral and entropy analysis \cite{Liang2010}, matching pursuit algorithm \cite{Sorensen2010}, tensor discriminant analysis \cite{Nasehi2013}, multifractal detrended fluctuation analysis \cite{Zhang2017}, and cross-bispectrum analysis \cite{Mahmoodian2019}.
		While KNN classifiers have been developed, among many others, with features from the fractal dimension \cite{Polychronaki2010}, and non-linear dimension reduction of frequency domain parameters \cite{Birjandtalab2017}. In addition to these two popular approaches, Acharya et al. have recently been the first to develop a seizure detection method using a convolutional neural network \cite{Acharya2017}.
	
	In general, these methods have good performance thanks to the advanced signal processing techniques they rely on. However, they consequently have high computational costs \cite{QuinteroRincon2018a}. In this work, we combine for the first time the two-point difference algorithm and the quadratic linear-parabolic model to design an original model-based statistical classification method to detect epileptic seizures in EEG. The proposed feature-vector is simple and fast to calculate from the quadratic model. Our classification method is based on bootstrap-aggregated (bagged) decision trees. This approach combines results from several decision trees to overcome the overfitting effect due to the variability in the training data \cite{Breiman2001}. Using the different subset of features (subspace sampling) to build the weak-tree classifiers reduces the training time \cite{Flach2012}, see \cite{Manzouri2018,Douget2017,Donos2015,Lopez2015} for some recent works in seizure classification using EEG.
	
	The remainder of the paper is organized as follows. Section \ref{sec:meth} describes the proposed methodology where the linear-parabolic curve fitting based on two-point central difference is introduced. In Section \ref{sec:res} the proposed methodology is applied to real EEG signals from patients suffering from epileptic seizures using the Children Hospital Boston database. Results of classification using the random forest  to detect seizure events are presented. Conclusions, limitations, advantages, and perspectives  for future work are finally reported in Section \ref{sec:con}.
	
	\section{Methodology}
	\label{sec:meth}
	
	Let $\bs X \in \mathbb{R}^{N\times M}$ denote the matrix of
	$M$ EEG signals $\boldsymbol{x}_m \in \mathbb{R}^{N\times 1}$ measured simultaneously on different channels and at $N$ discrete time instants. 
	
	The proposed methodology is composed of five stages. 
	The first stage divides the original signal $\bs X$ into a set of non-overlapping 1-second segments using a rectangular sliding window $\bs \Omega = \bs \Omega_0 \left( w-\frac{W-1}{2}\right)$ with $0 \leq w\leq W-1$, so that $ \bs X[n] = \bs \Omega[n] \bs X$. 
	In the second stage, the two-point central difference algorithm is used to estimate the coefficients $b[k]$ of the filter for each 1-second segment $\bs X[n]$. The signal is then convolved by the resulting filter $\widetilde{\bs X}[n]= \bs X[n] \ast b[k]$. The third stage fits a quadratic linear-parabolic model to the filtered signal, and estimates the associated parameters. 
		The fourth stage calculates four statistical parameters for each EEG segment, namely the weighted sum of squared residuals ($\zeta$), the R-square value ($\phi$), the adjusted R-square value ($\sigma$) and the root mean squared error ($\psi$), to assess the quadratic model fitting. Finally, in stage five, the feature-vector $\rho=[\zeta,\phi, \sigma,\psi]$ associated with each segment is classified using a random forest classifier in order to discriminate between seizure and non-seizure. 
	
	The following sections present the two-point central difference algorithm, the model fitting method based on curve fitting and the associated statistical parameters, and the random forest classifier.
	
	\subsection{Two-point central difference algorithm}
	\label{sec:centraldiff}
	
	The principle of the two-point central difference algorithm consists in subtracting non-adjacent, regularly spaced, pairs of points. The objective is to extract a slope from $\bs X[n]$: 
	\begin{align}
	\label{eq:centraldiff}
	X'[n]=\frac{\bs X[n+L]-\bs X[n-L]}{2LT_{s}},
	\end{align}
	where $L$ is called the skip factor that defines the distance between points, $T_{s}$ is the sample interval scaling. 
	Taking the $Z$-transform of both sides, we obtain
	\begin{align}
	\bs X'(z) = \frac{z^{L}-z^{-L}}{2LT_{s}}X(z).
	\label{eq:zeta}
	\end{align}
	The transfer function of \eqref{eq:zeta} is given by
	\begin{align}
	\bs Y(z,L)=\frac{\bs X'(z)}{\bs X(z)} = \frac{z^{L}-z^{-L}}{2LT_{s}}.
	\end{align}
	Thus, the frequency response is obtained by replacing $z$ by $e^{j\Omega}$
	\begin{align}
	\bs Y(e^{j\Omega},L) = \frac{e^{jL\Omega}-e^{-jL\Omega}}{2LTs} = \frac{jsin(L\Omega)}{LT_{s}}.
	\label{eq:response}
	\end{align}
	
	Using a symmetric FIR (finite impulse response) filter, the two-point central difference algorithm is based on an impulse function containing two coefficients of equal but opposite sign spaced $L$ points apart with the following coefficients
	\begin{align}
	\label{eq:fir}
	b[k]= \left \{ \begin{matrix} 
	\frac{1}{2LT_{s}} & k=-L \\ 
	-\frac{1}{2LT_{s}} & k=+L \\
	0 & k \neq \pm L. \\
	\end{matrix}\right. 
	\end{align}
	Note that, FIR filters are free from stability problems and they cause no time delay and no phase distortion within the pass-band.
	
	A convolution operation is estimated between each 1-second segment  $\bs X[n]$ and the coefficients $b[k]$ in order to obtain a new filtered signal in a bandwidth frequency until 50 Hz: 
	\begin{align}
	\widetilde{\bs X}[n] = \bs X[n] \ast b[k].
	\label{eq:conv}
	\end{align}
	Note that, this is the effective bandwidth of the filter where the EEG activity has an important clinical relevance \cite{Sanei2013}. This  allows to automatically reject the high line-noise artifacts greater than 50 Hz as $L=F_{s}/5$, as the sampling rate $F_{s}$ is the 256 Hz. Note that $L$ is rounded. 
	We refer the reader to \cite{Bahill1983,Tseng2008,BiosignalProcessing2014} for a comprehensive treatment of the mathematical properties of two-point central difference algorithm.
	
	\subsection{Quadratic linear-parabolic model}
	\label{sec:parabolic}
	We propose to fit a quadratic linear-parabolic model to the filtered signal. Precisely, the model is fitted to the couple $\left( \widetilde{\bs X}, \widetilde{\bs X}^{2}\right)$, where $\widetilde{\bs X}$ is the complete filtered signal, obtained by concatenating the segments given in equation   \eqref{eq:conv}. The idea is to fit the signal with a curve of the form:
	\begin{align}
	y= a \; sin(x-\pi) + b \; (x-10)^{2} + c.
	\label{eq:linear}
	\end{align}
	The fitting is performed using the least squares method \cite{LeastSquaresDF2013,Neuhauser2014,Samarasinghe2014} to estimate the three parameters $a,b,c$. 
	
	\subsection{Model-fitting statistical parameters}
	\label{sec:features}
	In order to assess the fitting of the curve given in equation \eqref{eq:linear}, the following four statistical parameters are estimated. Note that $y$ is the observed data, $\hat{y}$ is the predicted value using the quadratic linear-parabolic model, and $\bar{y}$ is the mean of the observed data.
	
	\emph{Weighted sum of squared residuals} ($\zeta$): 
	It is used to measure the total deviation of the response values from the predicted values, and is defined as
	\begin{align}
	\zeta=\sum_{i=1}^n w_i\left(y_i-\hat{y}_i\right)^{2}.
	\label{eq:sse}
	\end{align}
	The weights allow taking into consideration the different uncertainties of the measurements, and are calculated as follows 
	\begin{align}
	w_i =\frac{1}{\sigma_i^{2}},
	\label{eq:weights}
	\end{align}
	where $\sigma^{2}$ is the variance.\\
	\emph{R-square} ($\phi$): It is the square of the correlation between the response values and the predicted response values, and is defined as
	\begin{align}
	\phi= 1- \sum_{i=1}^n  \frac{\left(y_i-\hat{y}_i\right)^{2}}{\left(y_i-\bar{y}_i\right)^{2}}.
	\label{eq:rsquare}
	\end{align}\\
	\emph{Adjusted R-square} ($\sigma$): It adjusts the R-square residual degrees of freedom, and is defined as
	\begin{align}
	\label{eq:adjrsquare}
	\sigma &= 1- \frac{(1-\phi)(n-1)}{n-m-1},
	\end{align}
	where $n$ is the number of response values and $m$ is the  estimated fitted coefficients from the response values. A value of $\sigma$ closer to 1 indicates a better fit.\\
	\emph{Root mean squared error} ($\psi$): It is an estimate of the standard deviation of the random component in the data, and is defined as
	\begin{align}
	\psi = \sqrt{\frac{\zeta}{n-m}}.
	\label{eq:rmse}
	\end{align}
	
	\subsection{Random Forest classification}
	\label{sec:rforest}
	In this section, we present a classification method to identify seizures. The four statistical parameters presented above are used as classification features. A random forest classification technique is adopted. Random forest is an ensemble learning technique that combines the Bagging algorithm and the random subspace method using decision trees as the base classifier.
	The goal is to discriminate between seizure and non-seizure events, specifically the seizure onset through the feature predictor vector $\rho=[\zeta,\phi, \sigma,\psi]$ associated with each EEG segment $\widetilde{\bs X}$.
	The idea is to design a prediction function $f(\rho)$ to estimate the binary response for two classes, $\Gamma=\{\gamma_1,\gamma_0\}$, where $\gamma_1=1$ for a seizure event and $\gamma_0=0$ for a non-seizure event. 
	A loss function $L(\Gamma,f(\rho))$ determines the prediction function as follows
	\begin{align}
	E_{\rho,\Gamma} \left(  L(\Gamma,f(\rho)) \right),
	\end{align}
	where $E_{\rho,\Gamma}$ denotes the expectation with respect to an unknown joint distribution $P_{\rho,\Gamma}(\rho,\Gamma)$ of $\rho$ and $\Gamma$. For the classification, the next zero-one loss function is used 
	\begin{align}
	L(\Gamma,f(\rho)) =  I(\Gamma \neq f(\rho)) =
	\begin{cases}
	1 \text{ if }  &\Gamma =       f(\rho)  \approx \text{ for } \gamma_1 \text{ or seizure},\\
	0 \text{ if }  &\Gamma \neq  f(\rho)  \approx \text{ for } \gamma_o \text{ or non-seizure}.\\
	\end{cases}
	\end{align}
	The ensemble learning approach constructs $f$ by using $J$ decision-tree base-classifiers  
	\begin{align}
	\{h_1(\rho,\Theta_1), \cdots, h_J(\rho,\Theta_J)\}  
	\label{eq:base}
	\end{align}
	where $\Theta_j$, $j=1,\cdots,J$, is an independent collection of random variables.  These base classifiers are combined by majority vote, so that
	\begin{align}
	f(\rho) = \argmax_{\gamma \in \Lambda} \sum_{j=1}^J I(\gamma = h_j(\rho)). 
	\label{eq:pred}
	\end{align}
	
	Consider a dataset $\mathcal{D}=\{ d_1, d_2,..,d_N\}$ with $d_i=(\rho_i,c_i)$, where $c_i$ is a class label, $\rho$ denotes the vector predictor and $\gamma_i$ the response. For a particular realization $\theta_j$ of $\Theta_j$, the fitted tree is denoted $\hat{h}_j(\rho,\theta_j,\mathcal{D})$. The random forest classifier uses random subspace in two ways. First,  each tree is fitted to an independent bootstrap sample from the original data. The randomization involved in bootstrap sampling gives one part of $\Theta_j$. Second, the best split is retained for all $\rho$ predictors according to the randomly selected subset of $m$ predictor variables.
	
	The randomization used to sample the predictors gives the remaining part of $\Theta_j$. The resulting class predicted is by the majority combination of unweighted voting of the trees, see algorithm \ref{alg:rf}. We refer the reader to \cite{Breiman2001,ZhangMa2012,Flach2012} for a comprehensive treatment of the properties of random forest classifier.
	\vspace*{1.5em}

	\begin{algorithm}[!ht]
		\label{alg:rf}
		\KwData{data set $\mathcal{D}$; predictor vector $\rho$}
		\KwResult{ensemble of tree models whose predictions are to be combined by
			voting or averaging.}
		\For{$j=1$ to $J$}{
			\begin{algorithmic}[1] 
				\State Take a bootstrap sample $\mathcal{D}_j$ of size $N$ from $\mathcal{D}$.
				\State Using the bootstrap sample $\mathcal{D}_j$ as the training data, fit a tree using binary recursive partitioning:
				
				\hskip0.5em  $a$. Start with all observations in a single node \;
				\hskip0.5em  $b$. Repeat the following steps recursively for each unsplit node until the stopping criterion is met:\
				
				\hskip1.2em $i$. Select $m$ predictors at random from the $\rho$ available predictors. \;
				\hskip1.2em  $ii$. Find the best binary split among all binary splits on the $m$ predictors
				from Step i.\;
				\hskip1.2em  $iii$. Split the node into two descendant nodes using the split from Step ii.\;
			\end{algorithmic}
		}
		To make a prediction at a new point, use equation \eqref{eq:pred}.
	\vspace*{0.5em}
	\caption{Random forest classifier algorithm}
	\end{algorithm}

\newpage	
	\section{Experimentation and discussion}
	\label{sec:res}
	
	In this section, we evaluate the proposed methodology using the Children Hospital Boston database, which presented in the following section. 
	
	\subsection{Data}
	\label{sec:database}
	For the experimentation purpose, we considered  a dataset from the Children's Hospital Boston database \cite{Shoeb2004,Goldberger2000}, which consists of 22 EEG recordings from pediatric subjects with intractable seizures. All signals were sampled at 256 Hz with 16-bit resolution by using the International 10-20 system. The set of recordings lasted on average 35 minutes for 30 subjects in total; 2 hours for 4 subjects; and 12 hours for 2 other subjects. Taken together the recordings account for 60 hours of EEG recordings and 139 seizures. No distinctions regarding the types of seizure onsets were considered; the data contains focal, lateral, and generalized seizure onsets. Furthermore, the recordings were made in a routine clinical environment, so non-seizure activity and artifacts such as head/body movement, chewing, blinking, early stages of sleep, and electrode pops/movements are present in the data.  In this work, we have used 66 epochs from 9 different subjects, see Table \ref{tab:data}. Each recording contains a seizure event, whose onset time has been labeled by an expert neurologist. Moreover, for each seizure segment, the neurologist also selected one adjacent non-seizure signal segment before and after the seizure, of the same length to represent healthy brain activity. 
	In total, we considered 33 seizures and 33 non-seizure events. Non-seizure control events have been chosen just before seizures. This choice is justified for two reasons. First, we are interested in identifying seizure onset, therefore it is most important to distinguish its signal from preceding normal activity. Second, the EEG activity following a seizure remains chaotic for a long time and comparing it to seizure onset is not pertinent. The selected signals had the same montage by using 23 scalp EEG channels.

\begin{longtable}{||c|| c|| c|| c|| c||}
\hline\hline
Epoch	& Seizure & Duration (sec) & Genre & Age (years)\\ \hline \hline
\endfirsthead

\hline \hline
Epoch	& Seizure & Duration (sec) & Genre & Age (years)\\ 
\hline \hline
\endhead

\multicolumn{2}{c}{Continued on next page.}
\endfoot
\endlastfoot
	
01	& 	$01\_03$ 	& 40  & F & 11\\
02	& 	$01\_04$ 	& 27  & & \\
03	& 	$01\_15$  & 40  & & \\
04	& 	$01\_18$  & 90  & & \\ 
05 	& 	$01\_21$  & 115 & & \\
06	& 	$01\_26$  & 101 & & \\ 
\hline \hline
07	& 	$02\_16$  & 82  & M & 11\\
08 	& 	$02\_16$  & 21 & & \\
09 	& 	$02\_19$  & 9 & & \\
\hline \hline
10	& 	$03\_01$  & 52 & F & 14\\
11	& 	$03\_02$  & 55 & & \\
12	& 	$03\_03$  & 69 & & \\
13	& 	$03\_04$  & 52  & & \\
14	& 	$03\_34$  & 47  & & \\
15	& 	$03\_35$  & 64 & & \\
16	& 	$03\_36$  & 67 & & \\
\hline \hline
17	& 	$05\_17$  & 120 & F & 7\\
18	& 	$05\_22$  & 117 & & \\
\hline \hline
19	& 	$06\_18$  & 12  & F & 1.5\\
20	& 	$06\_24$  & 16  & & \\
\hline \hline
21	& 	$07\_12$  & 86 & F & 14.5\\
22	& 	$07\_13$  & 144  & & \\
23	& 	$07\_19$  & 83  & & \\
\hline \hline
24	& 	$08\_02$  & 189  & M & 3.5\\
25	& 	$08\_05$  & 190  & & \\ 
26	& 	$08\_21$  & 338  & & \\ 
\hline \hline
27	& 	$09\_06$  & 64  & F & 10\\ 
28	& 	$09\_19$  & 64  & & \\ 
\hline \hline
29	& 	$10\_20$  & 10 & M & 3\\
30	& 	$10\_27$  & 65 & &  \\ 
31	& 	$10\_30$  & 62 & &  \\ 
32	& 	$10\_31$  & 76 & &  \\ 
33	& 	$10\_89$  & 54 & &  \\
\hline \hline
\caption{Length of the 33 seizures used in this study. The duration has an average of 79,4 sec.}
\label{tab:data}
\end{longtable}

\subsection{Model fitting}
This section presents the results of the quadratic linear-parabolic model fitting.  Table \ref{tab:coeffs} presents the three coefficients $(a,b,c)$ estimated according to equation (\ref{eq:linear}). Coefficients for seizures and non-seizures are shown separately, for illustration purpose only. This allows us to note the significant difference between values corresponding to the two classes. The actual quadratic linear-parabolic model is fitted to the entire signal, without distinction of seizure and non-seizure. 

		\begin{table}[!ht]
			\centering
			\begin{tabular}{||c|| c|c|| c|c||}
				\hline \hline 
				& \multicolumn{2}{|c||}{Non-Seizure} & \multicolumn{2}{|c||}{Seizure}  \\ \hline \hline 
				Coefficients	& Value & CB95\% 			& Value & CB95\% 			\\ \hline \hline
				$a$ 	& -0.6855 & (-4.095, 2.725)  &  1.368  & (-6.717, 9.453)  \\ 
				$b$ 	& 0.9989  & (0.9989, 0.9989) &  0.9998 & (0.9998, 0.9998) \\ 
				$c$ 	& -36.05  & (-38.53, -33.57) 	 &  -35.45 & (-41.47, -29.43) \\ 
				\hline \hline 
			\end{tabular}
			\caption{The quadratic linear-parabolic model coefficients and the associated 95\% confidence bounds (CB).}
			\label{tab:coeffs}
		\end{table}
	
\begin{figure}[!ht]
		\centering
		\subfigure[Non-Seizures]{\includegraphics[width=58mm]{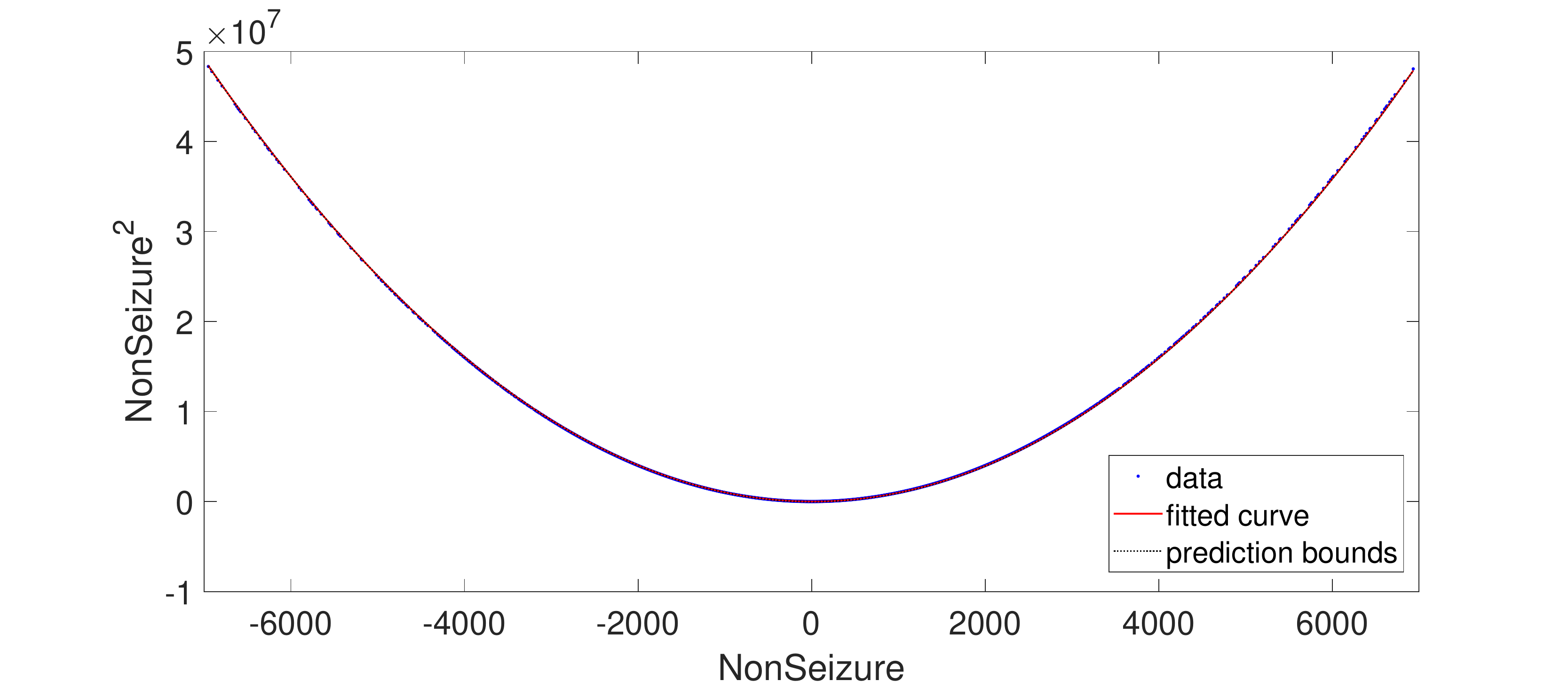}}
		\subfigure[Seizures]{\includegraphics[width=58mm]{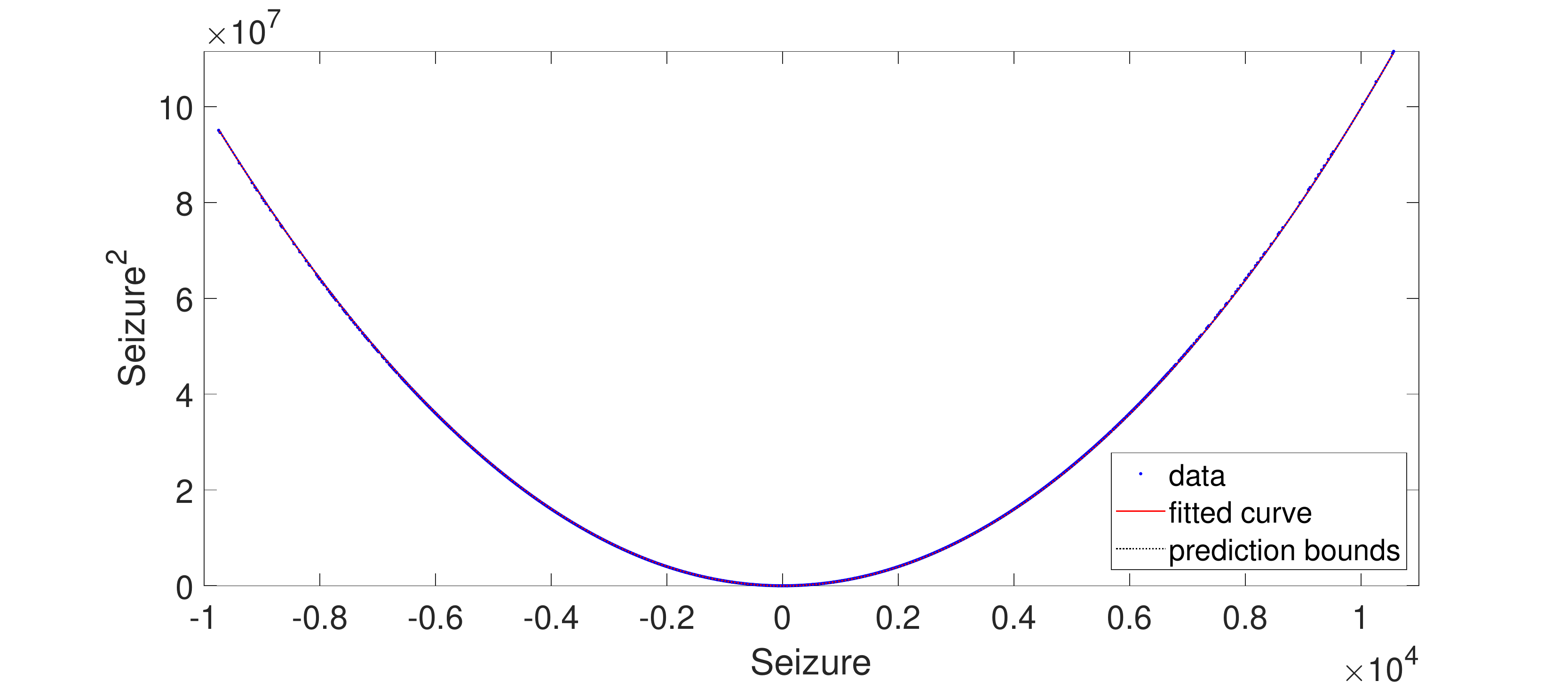}}
		\caption{Curves of the quadratic linear-parabolic model for seizure and non-seizure events. Note that the scale of the seizure curve is much larger than the one of the non-seizure curve. The x-axis represents $\widetilde{\bs X}$ while the y-axis represents $\widetilde{\bs X}^{2}$ (the unit on the X and Y axes correspond to the amplitude of the signal). The prediction bounds have a small uncertainty throughout the entire data range, therefore new observations can be predicted with high accuracy.}
		\label{fig:seizures}
	\end{figure}

Figure \ref{fig:seizures}.1 and \ref{fig:seizures}.2  show the quadratic linear-parabolic curves obtained for seizure and non-seizure segments, respectively. One notices that the proposed model \eqref{eq:linear} provides a visually precise fit to the EEG waveform. In addition, the prediction bounds have a small uncertainty throughout the entire data range, hence new observations should be predicted with high accuracy.
	
\subsection{Classification}
\begin{figure}[ht]
		\centering
		\includegraphics[width=1\columnwidth]{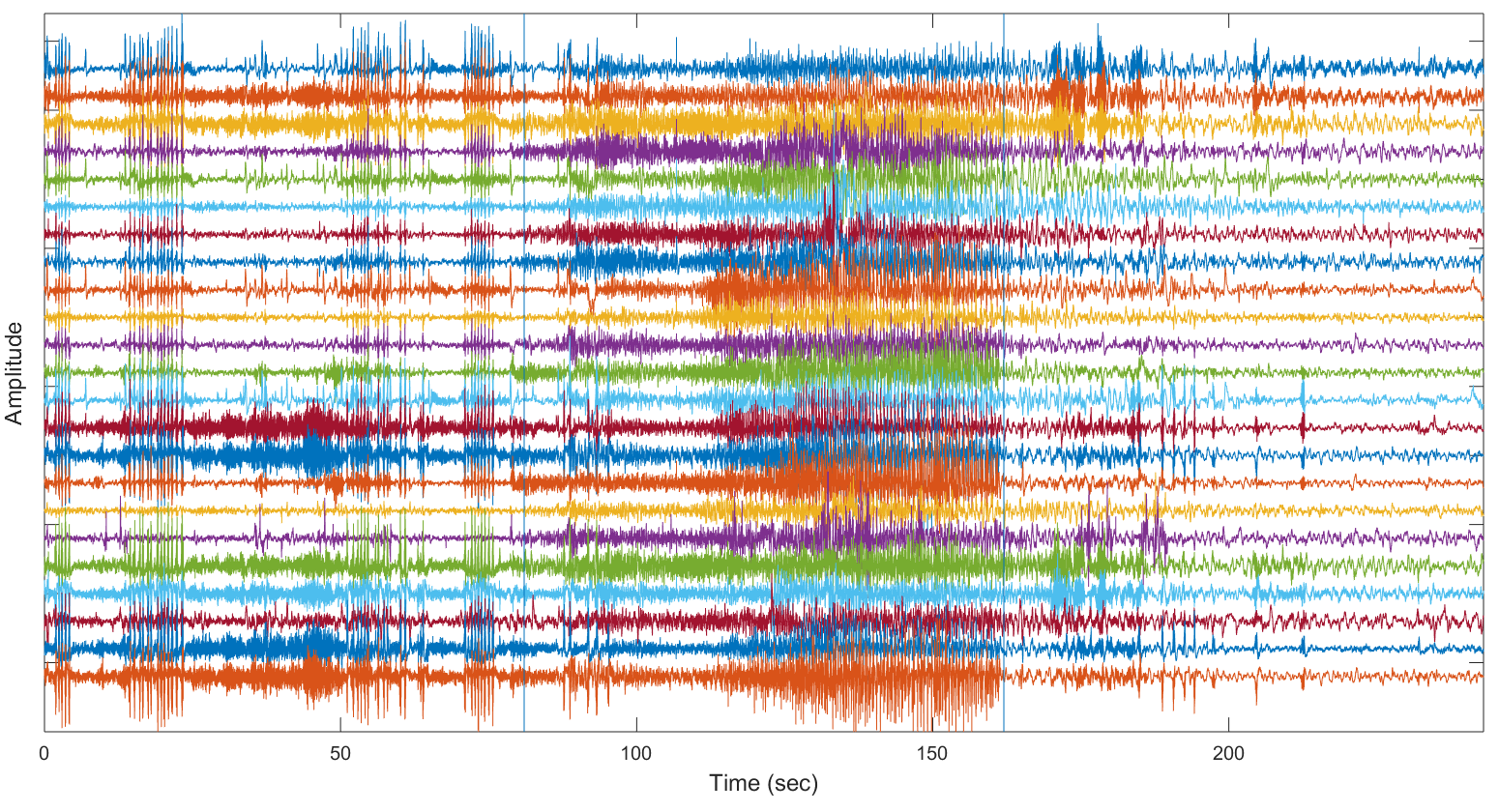}
		\caption{EEG raw example.  We can observe the great difficulty in discriminating between seizure and non-seizure. The onset detection begins in 81 sec according to the medical annotation but by non-expert visual inspection cannot reach the same conclusion. The y-axis scale for each channel is between $\pm 200$ mv.}
		\label{fig:scatterseizures}
	\end{figure}

The dataset described in section \ref{sec:database} has been used to train the classifier according to the proposed method. The capacity of the proposed classification scheme to discriminate between seizure and non-seizure events in order to detect the seizure onset in EEG signals has been assessed.  In Figure \ref{fig:scatterseizures}, one can notice the great difficulty in discriminating between seizure and non-seizure in EEG raw data.  The start and end of the seizure in this EEG signal were labeled by the neurologist using two lines. The first line divides the EEG signal at 81 sec (onset) and the second at 162 sec (offset).
	
Table \ref{tab:gof} reports the four statistical parameters, described in section \ref{sec:features}, calculated for all seizure and non-seizure segments. The weighted sum of squared residuals ($\zeta$) and the root mean squared error ($\psi$) show much larger values for seizure events with respect to non-seizure events, which  suggests that these features can be used to discriminate between seizure and non-seizures in EEG signals. While R-square ($\phi$) and adjusted R-square ($\sigma$) have values close to one for both types of events, which suggests that the model has high accuracy in the fit. This can be visually corroborated in Figure \ref{fig:seizures}.
	
	\begin{table}[!ht]
		\centering
		\begin{tabular}{||c|| c||c|| c||c||}
			\hline \hline 
			Events	& $\zeta$ & $\phi$ & $\sigma$ & $\psi$
			\\ \hline \hline
			Non-Seizure & 3.624e+14 & 0.9996 & 0.9996  & 4839 \\ 
			Seizure 	& 2.03e+15  & 0.9999 & 0.9999 & 1.145e+04 \\ 
			\hline \hline 
		\end{tabular}
		\caption{Goodness-of-fit statistics. The weighted sum of squared residuals  ($\zeta$) and root mean squared error ($\psi$) show much larger values for seizure events with respect to non-seizure events. While R-square ($\phi$) and adjusted R-squared ($\sigma$) have values near one for both, seizure and non-seizure events. These four parameters show good potential to discriminate between seizure and non-seizures in EEG signals.}
		\label{tab:gof}
	\end{table}
	
\begin{figure}[!ht]
		\centering
		\subfigure[Non-Seizures]{\includegraphics[width=58mm]{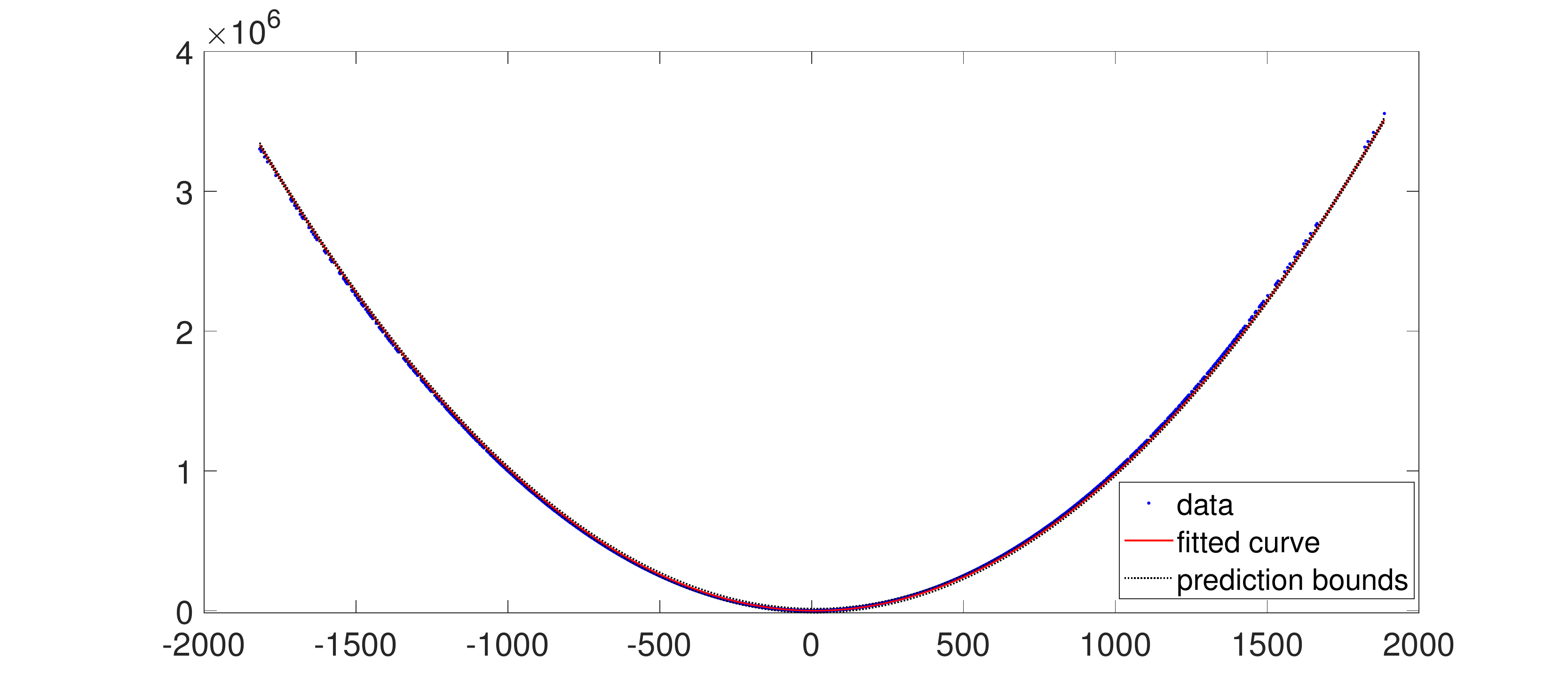}}
		\subfigure[Seizures]{\includegraphics[width=58mm]{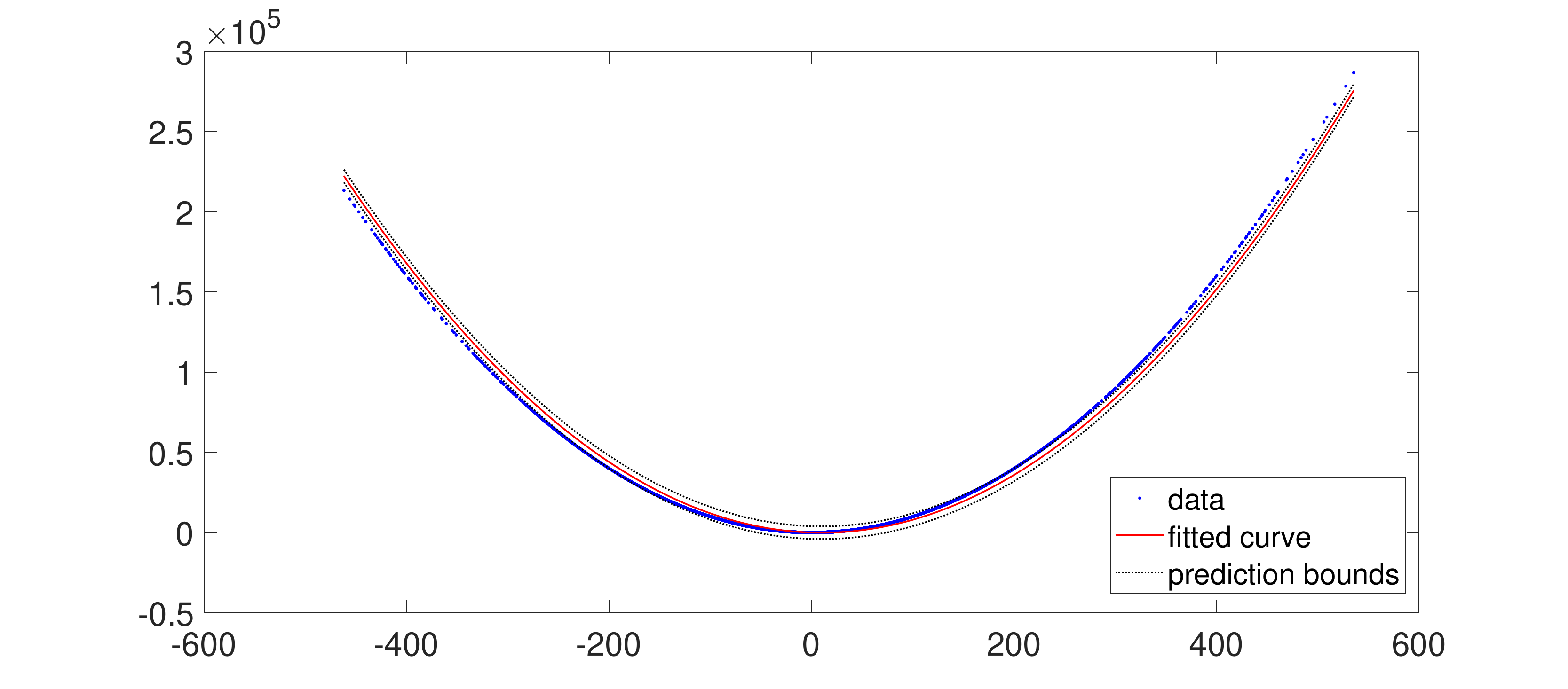}}
		\caption{Quadratic parabolic curve fitting example for one segment $\widetilde{\bs X}[n]$. We can see the waveform similarity with the previous figure \ref{fig:seizures}}
		\label{fig:seizuresExam}
	\end{figure}

Estimating the four statistical parameters at each 1-second segment allows us to detect changes between epileptic EEG seizures.  When we apply equation \eqref{eq:linear} to each 1-second segments, similar results are obtained as those applied to the signal as a whole. Figure \ref{fig:seizuresExam} shows the similarity with the previous figure \ref{fig:seizures}. Both seizure and non-seizure events have an excellent fitting with the linear-parabolic model, suggesting that the scale changes and the different values of the estimated parameters  can be used to differentiate  seizure and non-seizure events in EEG signals.
	
In order to cope with the high variability of EEG signals, the features were normalized to be in the range $[0,1]$. Table \ref{tab:gof2} reports the mean values for the four statistical parameters $\rho=[\zeta,\phi, \sigma,\psi]$ calculated for each 1-second  segment from $\bs X[n]$. The weighted sum of squared residuals ($\zeta$) has values closer to zero while R-square ($\phi$) and adjusted R-square ($\sigma$) have values closer to one. This shows that the model has a high fit accuracy and can be used for seizure detection, as corroborated by Figure \ref{fig:seizuresExam}.
	
	\begin{table}[!ht]
		\centering
		\begin{tabular}{||c|| c||c||c||c||}
			\hline \hline 
			Events	& $\zeta$ & $\phi$ & $\sigma$ &  $\psi$ \\ 
			\hline \hline
			Non-Seizure & 0.14  & 0.99 & 0.99 & 0.30 \\ 
			Seizure 	& 0.11  & 1.00 & 1.00 & 0.25 \\ 
			\hline \hline 
		\end{tabular}
		\caption{Means of the statistical parameters with values normalized between 0 and 1 for all each 1-second time segments. A value closer to zero in the weighted sum of squared residuals ($\zeta$) and Root mean squared error ($\psi$) and  values closer to one in R-square ($\phi$) and adjusted R-squared ($\sigma$) suggest a good fit quality and therefore can be used as a predictor.}
		\label{tab:gof2}
	\end{table}
	
\begin{figure}[!ht]
		\centering
		\subfigure[$\zeta$ vs. $\phi$]{\includegraphics[width=58mm]{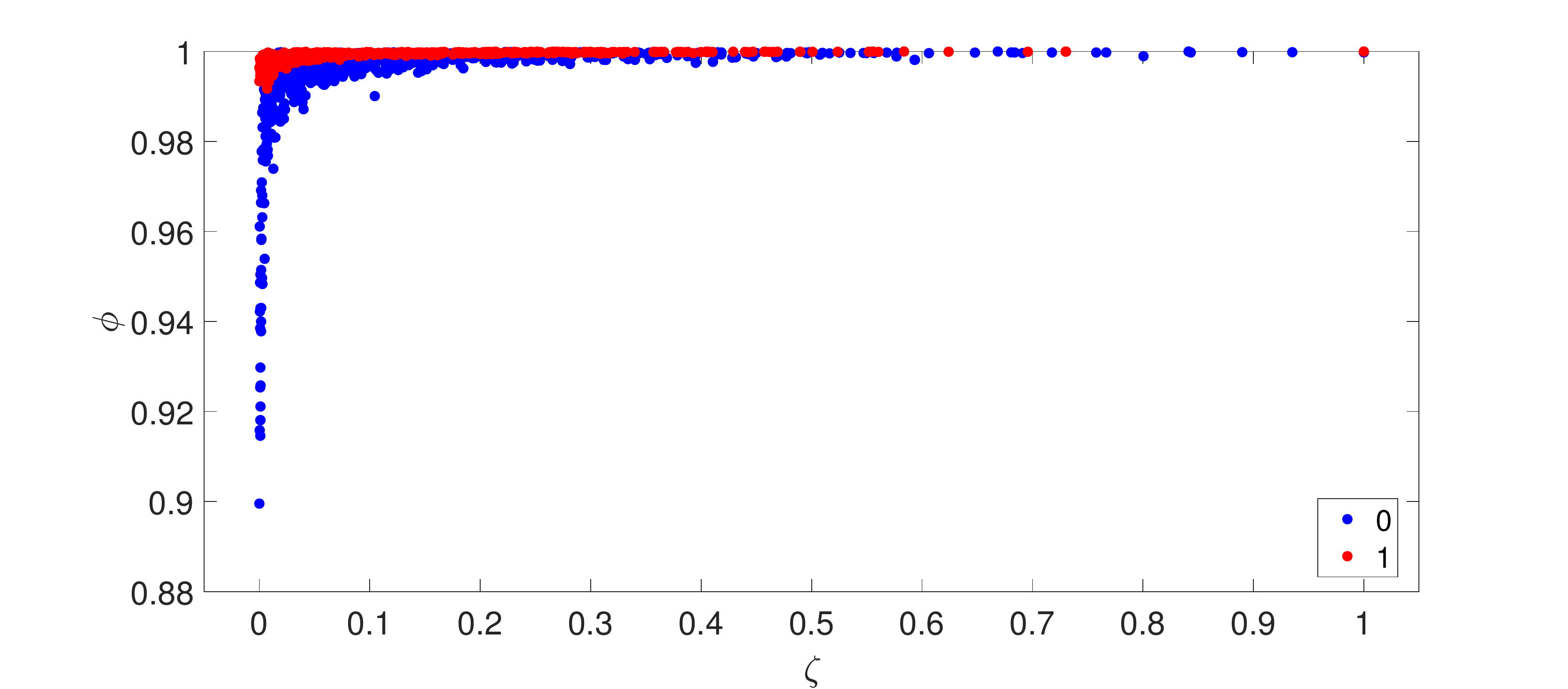}}
		\subfigure[$\zeta$ vs. $\psi$]{\includegraphics[width=58mm]{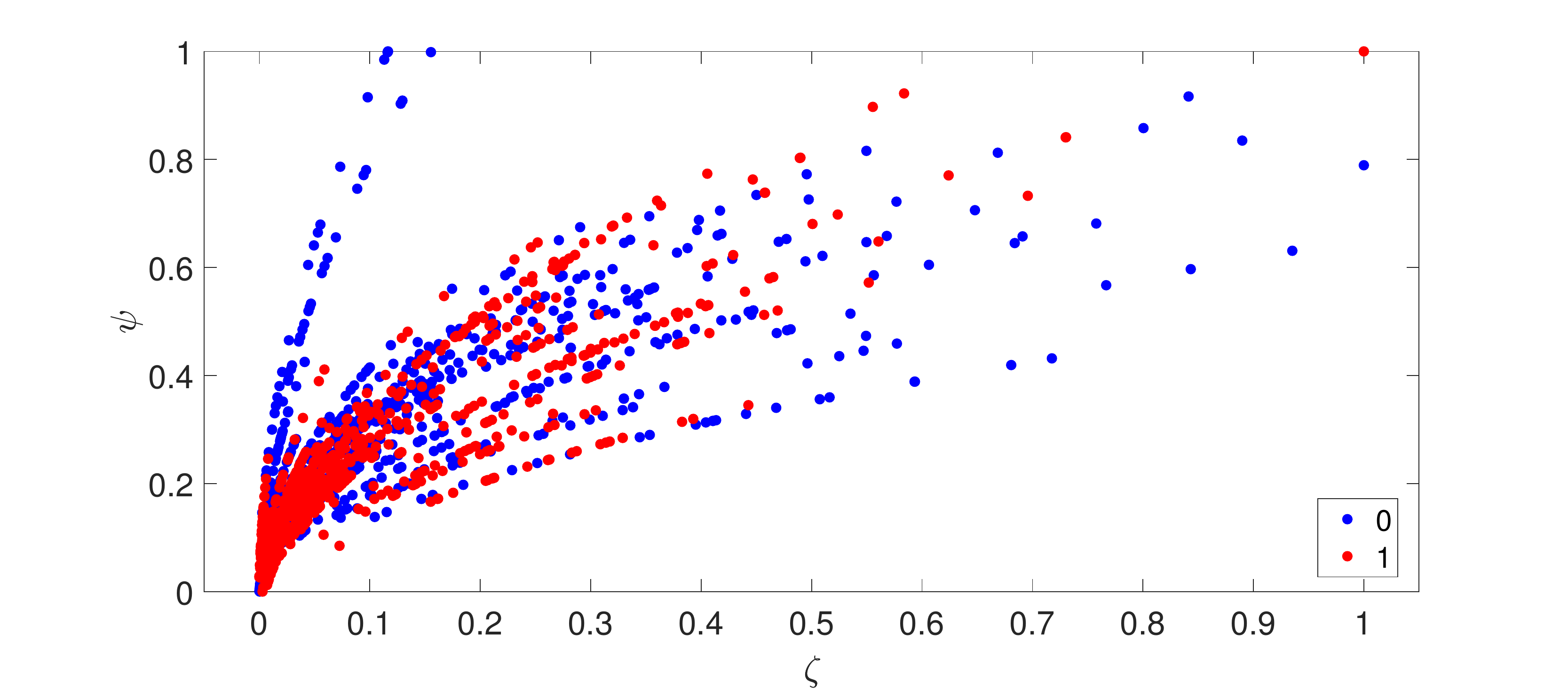}}
		\subfigure[$\psi$ vs. $\phi$ ]{\includegraphics[width=58mm]{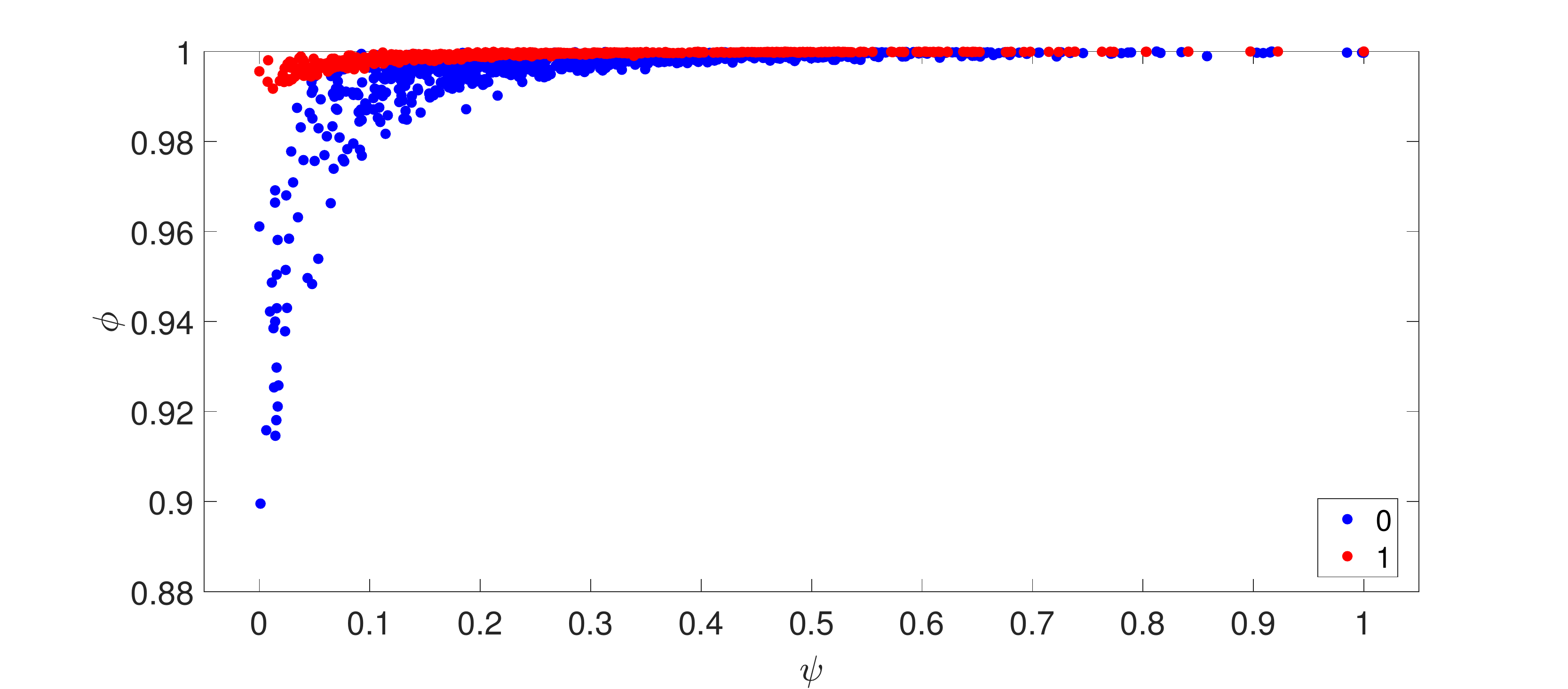}}
		\subfigure[$\phi$ vs. $\sigma$]{\includegraphics[width=58mm]{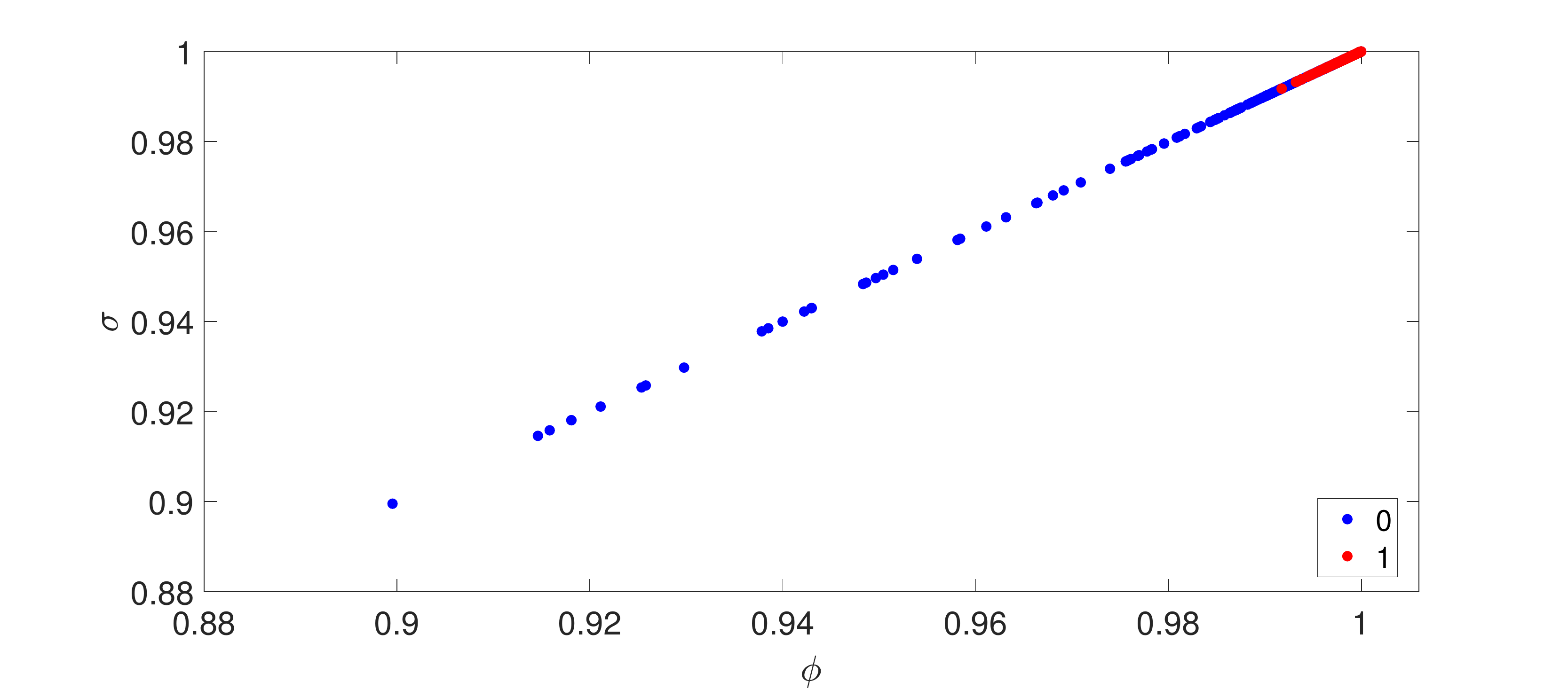}}
		\caption{Scatter plots for the four  statistical parameters $\rho=[\zeta,\phi, \sigma,\psi]$ observed through $1518$ events from 66 epochs, 33 seizures (red dots) and 33 non-seizures (blue dots) before the seizure.  The weighted sum of squared residuals ($\zeta$), R-square ($\phi$), adjusted R-squared ($\sigma$) and root mean squared error ($\psi$) shows a clear concentration of seizure events (red dots) with respect to non-seizure events (blue dots), which suggest that our model is potentially useful to detect between seizures and non-seizures events in EEG signals.}
		\label{fig:seizuresgof}
	\end{figure}

Figure \ref{fig:seizuresgof} shows the main scatter plots of couples of the normalized statistical parameters $(\zeta,\phi,\sigma,\psi)$. It is interesting to note that in $\zeta$ versus $\psi$ (fig.\ref{fig:seizuresgof}.a) and  $\psi$ versus $\phi$ (fig.\ref{fig:seizuresgof}.c) the seizure events (red dots) are concentrated closer to one with respect to non-seizure events (blue dots). In (fig.\ref{fig:seizuresgof}.b) $\zeta$ versus $\psi$ has a big tendency to go to the right in the seizure events (red dots) with respect to non-seizure events (blue dots). In (fig.\ref{fig:seizuresgof}.d) $\phi$ versus $\sigma$ shows a clear concentration of seizure events (red dots) with respect to non-seizure events (blue dots) on the upper right side. These scatter plots show that our predictor-vector $\rho=[\zeta,\phi, \sigma,\psi]$  is potentially useful to discriminate between seizures and non-seizures events in EEG signals.
	
To assess the performance of the proposed method, we adopted a supervised testing approach and used $1518$ events (1-second signals) to train and test the method with the 20-fold cross-validation technique of the predictor vector $\rho=[\zeta,\phi, \sigma,\psi]$. As explained in section \ref{sec:database}, the events are extracted from 33 seizure epochs and 33 non-seizure epochs, giving a total of 66 epochs from 9 different subjects. The method gives good classification performance, with 92\% true positives rate (TPR) or sensitivity, 96\% true negative rate (TNR) or specificity, 4\% false positive rate and 94.1\% accuracy. Please note that segments for training and testing are drawn randomly without consideration of their epochs. This process is repeated 1000 times to ensure not-bias in the partitioning.
	
\section{Conclusion}
\label{sec:con}
This work presented a new method to detect epileptic seizures in EEG signals. The proposed model-based classification method relies on the design of a specific filter using the two-point central difference algorithm. A linear-parabolic model is fitted to the filtered signal using mean squares. Four statistical parameters associated with the model fitting are used as classification features to discriminate seizure and non-seizure events. A random forest classifier has been adopted to evaluate the ability of these parameters to detect seizures. The proposed methodology was applied on $66$ balanced events from the Children Hospital Boston database. Results suggest that the proposed algorithm is a powerful tool for detecting epileptic seizure events in EEG signals achieving a 94.1\% accuracy.
	
The proposed method has two main advantages compared to existing methods. First, its low computational complexity makes it implementable in quasi-real-time. Second, the method can be easily extended to work separately with different brain rhythms, by adapting the skip factor $L$. Despite these advantages, the method suffers some limitations, mainly noise and robustness. 

The signal is affected by noise and artifacts due to the acquisition and pre-processing. This affects the precision of the results, especially when the method is generalized to complex epileptic forms. Outliers are not directly considered in the estimation procedure that needs more robustness.

Future work will focus on robust approaches to consider noise and artifacts. Optimization techniques will be investigated in order to remove outliers and improve the accuracy of the detection.
	
\bibliographystyle{unsrt}

\end{document}